\documentclass[aps,prl,twocolumn,reprint]{revtex4-1}
\usepackage{graphics}
\usepackage{epstopdf}
\usepackage{epsfig}
\usepackage{graphicx}
\usepackage{epsf,epic}
\usepackage{color}
\usepackage{subfig}
\usepackage{amsmath}
\usepackage{booktabs}
\usepackage{multirow}
\usepackage{physics}
\usepackage{hyperref}
\usepackage{amsfonts}
\usepackage{wrapfig}
\usepackage{breqn}
\usepackage{pstricks}
\usepackage{multirow}
\usepackage{fancyref}
\usepackage{pst-node}
\usepackage{float}
\usepackage{bm}
\usepackage{dcolumn}
\newcommand{\etal}{\textit{et al.\ }}
\newcommand{\ie}{\textit{i.e.\ }}

\makeatletter
\usepackage{etoolbox} 
\appto{\appendix}{%
	\@ifstar{\def\theequation@prefix{A.}}%
	{}%
}
\preto\maketitle{%
  \begingroup\lccode`~=`,
  \lowercase{\endgroup
  \let\saved@breqn@active@comma~
  \let~}\active@comma 
}
\appto\maketitle{%
  \begingroup\lccode`~=`,
  \lowercase{\endgroup
  \let~}\saved@breqn@active@comma 
}
\makeatother
\begin{document}
\title{Topological quantum switch and controllable quasi 1D wires in antimonene}
\author{\href{https://www.santoshkumarradha.me}{Santosh Kumar Radha}}
\email{Corresponding author:srr70@case.edu}
\author{Walter R. L. Lambrecht}
\email{walter.lambrecht@case.edu}
\affiliation{Department of Physics, Case Western Reserve University, 10900 Euclid Avenue, Cleveland, OH-44106-7079}
\begin{abstract}
  Based on the recently found non-trivial topology of buckled antimonene,
  we propose  the conceptual design of a quantized switch that is protected by topology and a mechanism to create configurable 1D wire channels.
  We show that the topologically required edge states in this system can be turned on and off by breaking the inversion symmetry (reducing the symmetry
  from $S_6$ to $C_3$), which can be achieved by gating the system. This
  is shown to create a field-effect quantum switch projected by topology.
  Secondly we show that by locally gating the system with different polarity
  in different areas, a soliton-like domain wall is created at their
  interface, which hosts  a protected electronic state, in which
  transport could be accessed by gated doping. 
  \end{abstract}
\maketitle

\paragraph*{Introduction:} Mono-layer antimony (antimonene) in the honeycomb
lattice ($\beta$-Sb) is emerging rapidly as a 2D system exhibiting
various interesting physical phenomena.
Recently\cite{radha2019topological}
we showed that this system in its flat state (which is a nodal line
semimetal) is the first example of planar \textit{goniopolarity}: \ie
the transport carrier is  hole- or electron-like depending  on
in-plane direction.  We also showed that the much sought after Dirac cone annihilation physics can in principle be achieved in this system by controlling
the in-plane tensile strain. Two topologial transitions occur under
increasing buckling (decreasing in plane tensile strain) at which
pairs of Dirac cones (of opposite winding number) protected by
the $C_2$ crystal symmetry axes annihilate when  band inversion occurs between
bands beloging to different irreducible representations. 
In another  paper\cite{radha2020buckled}  we showed that the fully
buckled equilibrium structure of the free standing monolayer  is a
Higher Order Topological Insulator (HOTI) with protected corner states.
Chiel \etal\ \cite{chiel2019symmetry} showed that the buckled form could host mechanical edge modes leading to strain solitons.
It is particularly promising  that these need not be confined to
theoretical  predictions because
several experimental routes to achieve monolayers of $\beta$-Sb have
already been reported, including mechanical exfoliation and epitaxial
growth.\cite{Ares18review,JiJiangping16}

\begin{figure}[h]
  \includegraphics[width=\linewidth]{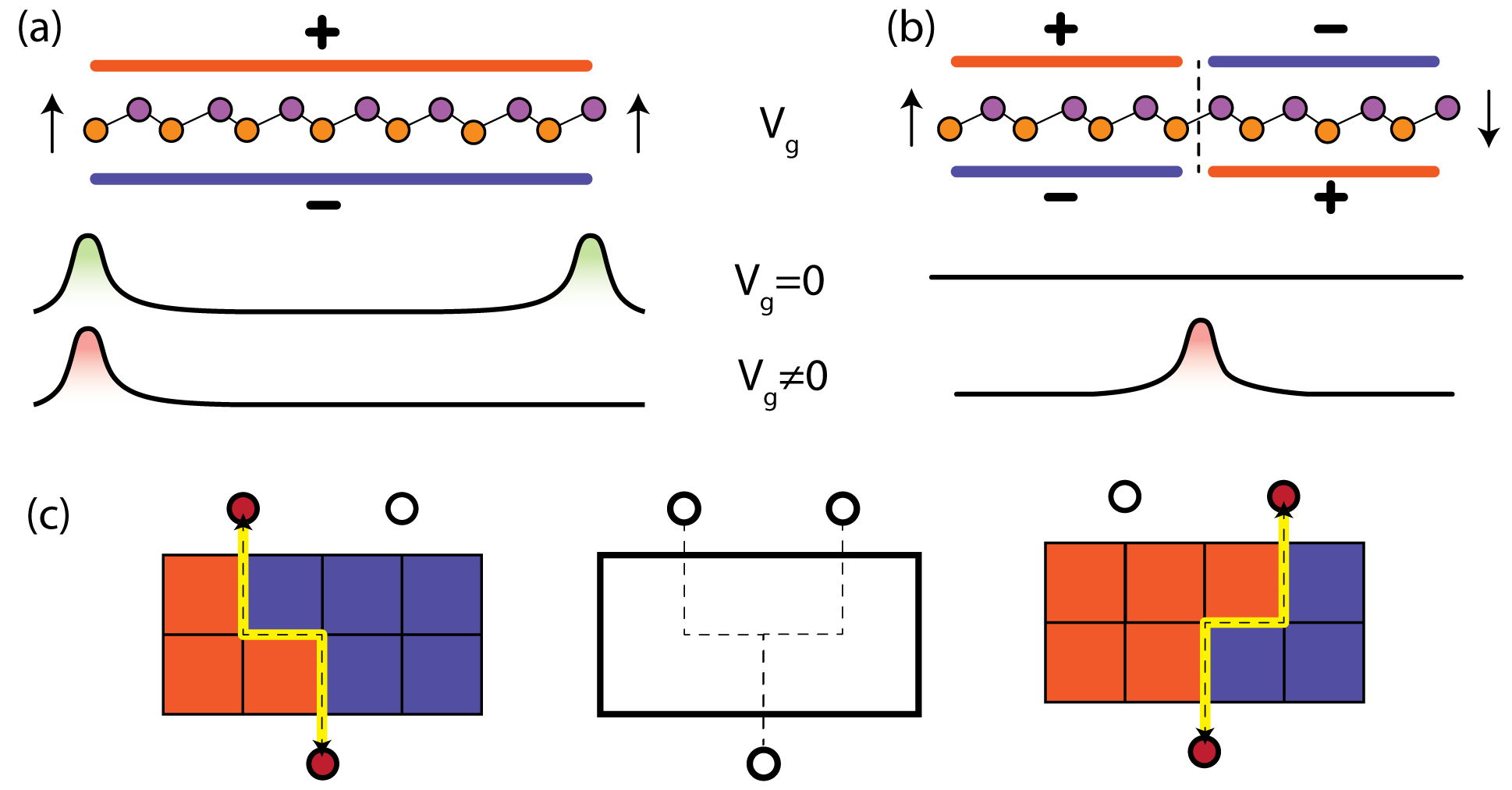} 
  \caption{Illustration of the two devices \textit{(a)} topological quantum switch \textit{(b)}quasi  1D domain wall channels \textit{(c)} Configurable 1D wire in a pixilated local gated system\label{fig:illustration}}. 
\end{figure}

Field effect transistors (FET) are the key elements in modern
electronics  and function by turning on and off the conductivity
through a channel between source and drain by means of a gate.
Typically, the channel is a 2D electron gas between two semiconductors
and the gate voltage is applied
through an insulating layer, such as an oxide in a MOSFET, which
thereby controls the carrier concentration in the 2D gas.
An important quantity is the ratio between on and off state resistance.
Here we propose a switch based on turning off the conduction
in topological edge states of a 2D topological crystal insulator
by means of an electric field gate. In the 2D Sb system
under consideration, the conducting edge states are topologically protected
by inversion symmetry and an electric field breaks
the inversion symmetry thereby turning the conducting channel off
in an abrupt way. Furthermore the conductance  itself can be
in the ballistic quantized regime.
This is illustrated in \autoref{fig:illustration} \textit{(left)}.

A somewhat related but  not identical  idea for a topology based 
switching  device  was proposed by Liu \etal\cite{JunweiLiu}. However
that device is based on breaking the mirror plane
symmetry, which protects a topological state, by means of an electric
field in the  3D material  SnTe. Strong effects on magnetotransport
due to mirror symmetry breaking were also reported by Wei \etal\cite{Wei18}.

A second intriguing idea in advanced electronics
is that of reconfigurable circuits. 
Cheng \etal\cite{Cheng11,Cheng13,Bi10} demonstrated this idea by creating a conducting local 
2D electron gas at the LAO/STO (LaAlO$_3$/SrTiO$_3$) interface induced
by removing surface adsorbed OH groups using a scanning tunneling microscope
tip along selected lines drawn on the surface of LAO a few layers above
the LAO/STO interface.  Here we propose  a different mechanism for creating
reconfigurable quasi 1D semi-conducting
channels in the monolayer Sb system. Furthermore the conductance in this 1D
system could show quantized conductance
in the low-temperature ballistic regime.\cite{Imryt87}
This is also based on breaking the inversion symmetry.
But instead of breaking it in the entire system, we show that by
locally breaking inversion with opposite polarity
one can form domain walls hosting localized 1D dispersing states in the gap
which can become semi-conducting channels for electron or holes. 
Unlike the drawing mechanism used in the LAO/STO system, which requires
nanoscale scanning probe type manipulations, the process
to reconfigure the path in our proposed scheme is much simpler.
First, it is a low energy process as one needs a very small amount of
bias to open and close the channel because its existence is based
on topology. Second it would only involve applying bias voltages
to a static configuration of localized pixelated gates 
without the need for mechanical motion. 
A schematic illustration of this device is shown in \autoref{fig:illustration} \textit{(right)}

The main idea behind both types of devices presented here is based on
applying an electric field normal to the layers which breaks
the inversion symmetry between the two sublattices because of the
alread present buckling which exposes each sublattice to a different
potential in the presence of an electric field. The use of an electric
field normal to the layers of a 2D system
to tune the electronic structure has been considered  before,
in particular for flat monolayer Sb.\cite{Hsu2016} However, in that case
the electric field's role is to break the horizontal mirror plane rather
than the inversion symmetry. The topological edge states in that
regime are different and result from the spin-orbit coupling. 

\paragraph*{Symmetry and topology:}
In monolayer honeycomb group-V systems, unlike in graphene,
the atomic $p$-orbital
derived bands are disentangled from the $s$-orbitals.\cite{radha2019topological}
In the completely flat limit, interaction between $p_z$ and $p_x,p_y$ is
absent because of mirror symmetry. Buckling breaks the mirror symmetry
and becomes a tuning parameter for the interaction and after a critical
angle ($\approx 33^\circ$ for Sb) or equivalently reduction of tensile in-plane strain, the system gaps up and attains the lowest energy stable structure
that has been synthesized\cite{JiJiangping16}. In this structure, all
three $\{p_x,p_y,p_z\}$ orbitals are hybridized  and determine the low-energy
physics near the gap, by forming a set of bonding and antibonding combinations
in the valence and conduction band respectively.
As described in Ref. \onlinecite{radha2020buckled} the valence band manifold
has bond-centered Wannier functions obeying $S_6$ symmetry. This
places the system in the ``obstructed atomic limit'' (OAL) as defined by
Bradlyn \etal\cite{Bradlyn2017,Cano18} which has non-trivial
crystalline topological (TCI)
properties, leading to weakly protected edge-states
as well as corner states. In Ref. \onlinecite{radha2020buckled}
we showed the system can be mapped to the bond-alternating honeycomb
or Kekul\'e lattice, which can be viewed as a 2D generalization of
the iconical Su-Schrieffer-Heeger (SSH) model whose alternating weak and
strong bonds  are well known to lead to protected end states
when the unit cell is chosen so that
the strong bonds are broken at the ends of a finite chain.
The $S_6$ group has 3D inversion $\cal{I}$ as well as $C_3$ symmetry
operations and it is the inversion symmetry which protects the
edge and corner states.\cite{Benalcazar19,Schindler19,radha2020buckled}
This inversion is directly connected with the identical Sb atoms in this
bipartite system, which leads to mid-gap 1D edge bands and corner states of 0D
symmetry preserving flakes.

Unlike for a  flat honeycomb lattice,  because of buckling,
one can easily break this
sub-lattice (inversion) symmetry by an applied electric field (or gating)
in the direction perpendicular to the layers, thereby creating a modulated on-site potential. This is obvious from \autoref{fig:illustration}.
This then leads to a $C_3$ point group as the reduced symmetry.
As a result the edge states gap up
and hence the fractionalized
electrons previously spread over
the two identical edges will now reside on the one edge (corner) with
the lower energy. This, in principle, leads from partially filled edge (corner)
states on both edges  to a completely filled and empty edge (corner) state
on opposite sides and hence a metal-insulator transition in the case
of dispersing edge states or a switch in which half of the corner states
switch from half occupied to fully occupied and the other ones to empty as
detailed next. 


\begin{figure}[!htb]
  \includegraphics[width=\linewidth]{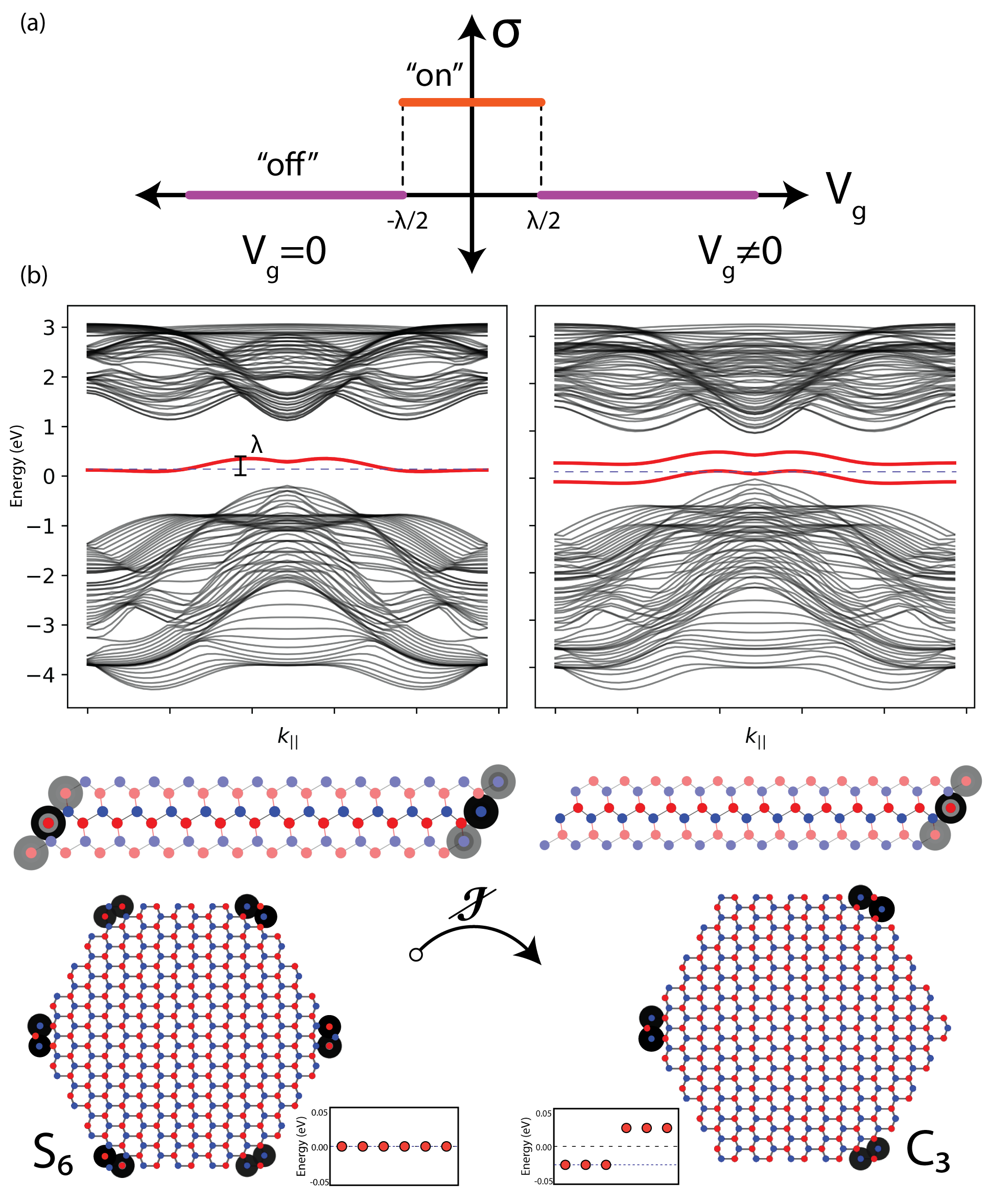} 
  \caption{ \textit{(a)} Operation of ``on'' and ``off'' state using the device \textit{(b)} band structure with gate voltage turned on and off and currosponding $\abs{\Psi}^2$ \textit{(c)} Effect of Inversion breaking in corner states of 0D system\label{fig:switch}}. 
\end{figure}

\paragraph*{Topological switch:} At low temperature, graphene is known
to have quantized low bias conductance in 1D armchair (zigzag) edge modes
given by\cite{PhysRevB.73.195411} $G=4n\frac{e^{2}}{h}\tilde{t}$
($G=2(2n+1)\frac{e^{2}}{h}\tilde{t}$)
with $\tilde t$ the transmission coefficients to the leads, whereby
the conductance occurs in steps but which are not exact
multiples of the quantum of conductance. 
In this nanoscale ballistic transport regime, the conductance is
determined by the number of 1D transverse modes contributing to the
transport.\cite{Imryt87}
In the present case of a 1D Sb nanoribbon and for zero bias normal to the
plane of the ribbon, 
we have two zero energy edge modes (one for each edge) with half occupation
for each spin channel, therefore 
giving $G=4\frac{e^{2}}{h}\tilde{t}$ conductance, when the Fermi level
lies within this band.  In contrast to graphene, where the edge bands connect
the host valence and conduction bands, 
the large gap in the 2D Sb host system ($\sim$ 3eV)
implies  a huge barrier before carriers can access the high/low lying
2D host bands in the ribbon giving us zero
conductance for any  Fermi level location in the gap
outside the  range $(E_{min},E_{max})$ set by the 1D dispersion of the
edge bands. Note that the dispersive nature of the 1D edge bands and their
bandwidth stem from the next nearest neighbor
interaction along the parallel direction and are hence small (of order 0.3 eV).
Breaking the inversion symmetry between the sublattices by a gate voltage,
as already described in the previous section, now opens a gap between the
edge bands as soon as the
splitting becomes larger than the 1D edge band dispersion.  
Assuming that the gate voltage
changes the on-site potentials on the two sub-lattices by
$\pm\Delta/2$ and thus opens a splitting $\Delta$ between the energy centers
of the two edge bands and a dispersion in the edge modes
of width $\lambda$, a gap between the
two edge bands occurs when $\abs{\Delta} > \lambda$ and one
enters the insulating limit with zero
conductance. Because $\lambda$ is small compared to the gap of the 2D system,
only a small gate voltage will be required to switch the device from its
``on'' (conducting) to its ``off'' or insulating state. 

There are three major advantages to device of this type compared to traditional
switches: (i) this switch has topological protection as the \textit{on} state is protected from perturbations including defects in the lattice because
of topological non-triviality (as shown in \cite{radha2020buckled});
(ii)  unlike traditional switches, we have quantized binary states; and
finally (iii) the barrier to turn off the switch requires only  low power
with a value set by the dispersion of the edge modes.

Another possible device that uses the inversion symmetry as switch is to use the same mechanism in the corresponding finite size symmetric hexagonal flakes of the system. 
With inversion symmetry preserving the $S_6$, one is guaranteed (spinless)
charge fractionalization of $e/2$ in each $\pi/6$ sector. But as soon as one
reduces the$S_6$ symmetry to $C_3$, the edge modes can interact and open
a gap. This leads to the $e/2$ charges being transferred to the respective previously inversion symmetric point.
(Including spin the charge trasfer occurs for each spin so becomes $e$.) Thus at zero bias voltage, where inversion symmetry is preserved, one has no potential difference between each $\pi/6$ sectors while a inversion breaking non-zero bias potential difference between $C_3$ symmetric inversion points as shown in \autoref{fig:switch}(c).

\paragraph*{1D Quantum wires:} We now look at selectively breaking inversion symmetry in the system with opposite polarity which we will show leads to a
domain boundary localized state in the gap. 
To gain qualitative insight, note that 
breaking inversion by biasing the system essentially acts like
adding a mass gap of $m\sigma_z$ to the 1D SSH system.
This breaks the chiral/inversion symmetry (in the 1D case) which can be easily seen in the finite 1D chain with one end being A $(+m)$ site and the other end being B $(-m)$ site. Thus the previously degenerate edge modes trivially gap with the occupied state being at the low energy $B$ side. The same physics holds here with one edge being occupied by the low potential part of the inversion broken sub-lattice and other side being empty when an inversion breaking bias is applied.

\begin{figure}[!htb]
  \includegraphics[width=\linewidth]{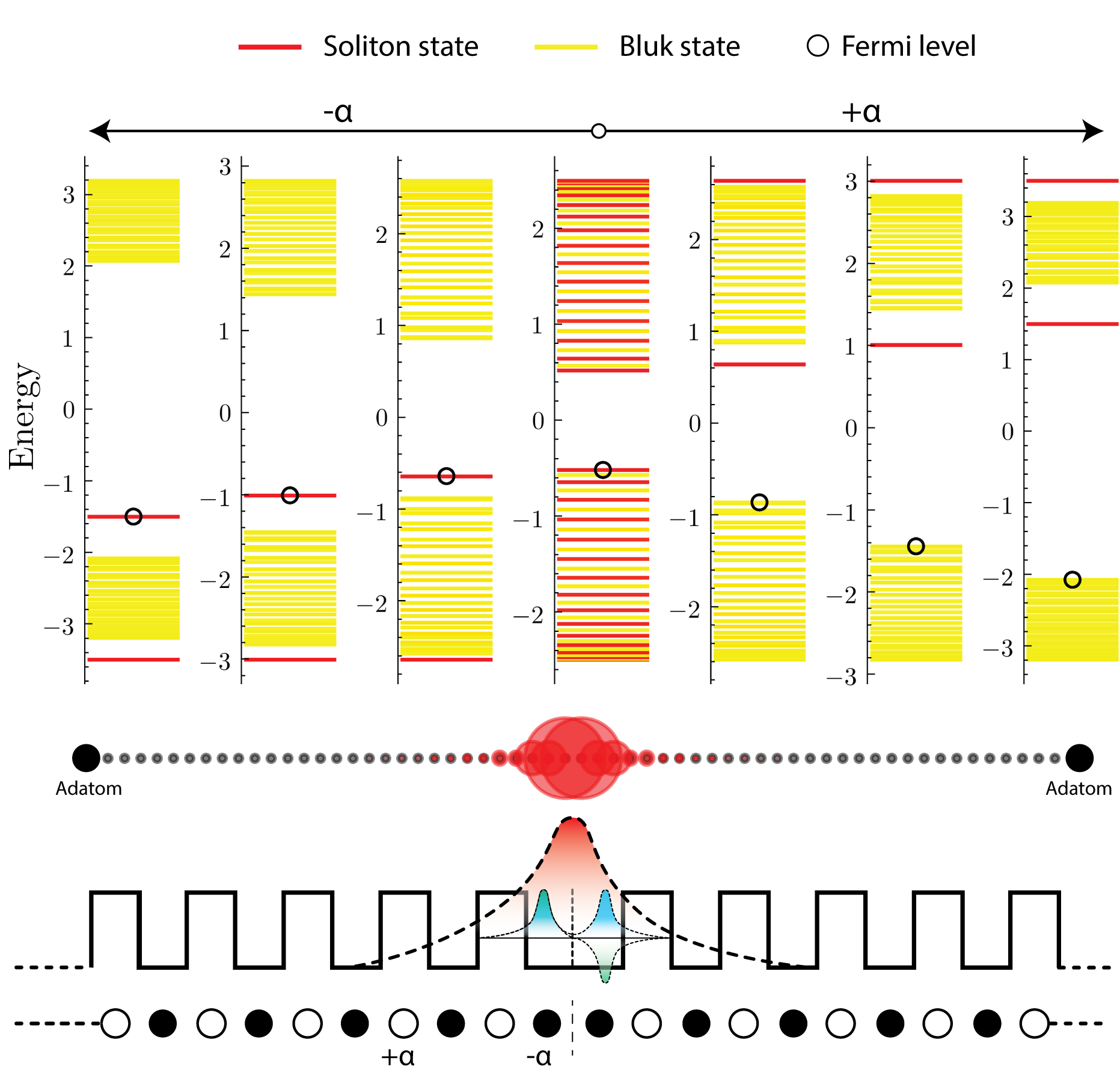} 
  \caption{ \textit{(top) } Energy spectrum of \eqref{eq:1} as function of $\alpha$ with $\delta t\neq0$ red levels are projected onto two sites near origin and yellow on bulk. (Energy scales are different)\textit{(middle)} $\abs*{\Psi}^2$ of the localized state. \textit{(bottom)} potential profile and the illustration of localized state and corresponding symmetric and antisymmetric wave-function \label{fig:SSH-soliton}}. 
\end{figure}

For creating a 1D quantum wire, we split the device into two regions, where in one region we apply a bias voltage of $+V$ and in the other $-V$.
This has the effect of creating alternating onsite terms on one side and
the opposite onsite terms on the other side with the junction between them having a domain wall of two $V$.
To understand the energetics of this system, we start by looking at the same effect in a simpler 1D SSH, which is nothing but the 1D projection of the
buckled Sb  2D system. The SSH Hamiltonian can be written as
\begin{dmath}
H(\delta t,\alpha)=\sum_i (t+\delta t) c_{i, A}^{\dagger} c_{i,B}+(t-\delta t) c_{i+1,A}^{\dagger} c_{i,B}+\left(-1^{\theta[i]}\right)\alpha \sum_i(c^\dagger_{i,A}c_{i,A}-c^\dagger_{i,B}c_{i,B})+h.c. \label{eq:1}
\end{dmath} 
where $c^{(\dagger)}_{i,j}$ creates (annihilates) an electron in unit cell $i$
on sublattice site $j$ and $\theta[x]$ is the Heaviside step function.
The first terms is the usual SSH Hamiltonian with $\delta t$ controlling the hopping anisotropy. For $\delta t=0$, we are at the critical point between
the trivial/non-trivial  regimes of the
SSH model and the system would be metallic with zero gap. 
However, the effect
we are concerned with here would still be valid because it depends
on the second term. We just want a gap to more clearly see the
edge states but whether the SSH gap is of the trivial or non-trivial
type is irrelevant for the following discussion. 
The second term here adds a mass gap of $\alpha \sigma_z$ for $i<0$ and $-\alpha \sigma_z$ for $i>0$. It is easy to see that this system has inversion {\sl w.r.t.} origin, the sites to the left and right of origin (A and B) gain the same value of $\alpha$ to their on-site term.

\begin{figure*}[!htb]
  \includegraphics[width=\linewidth]{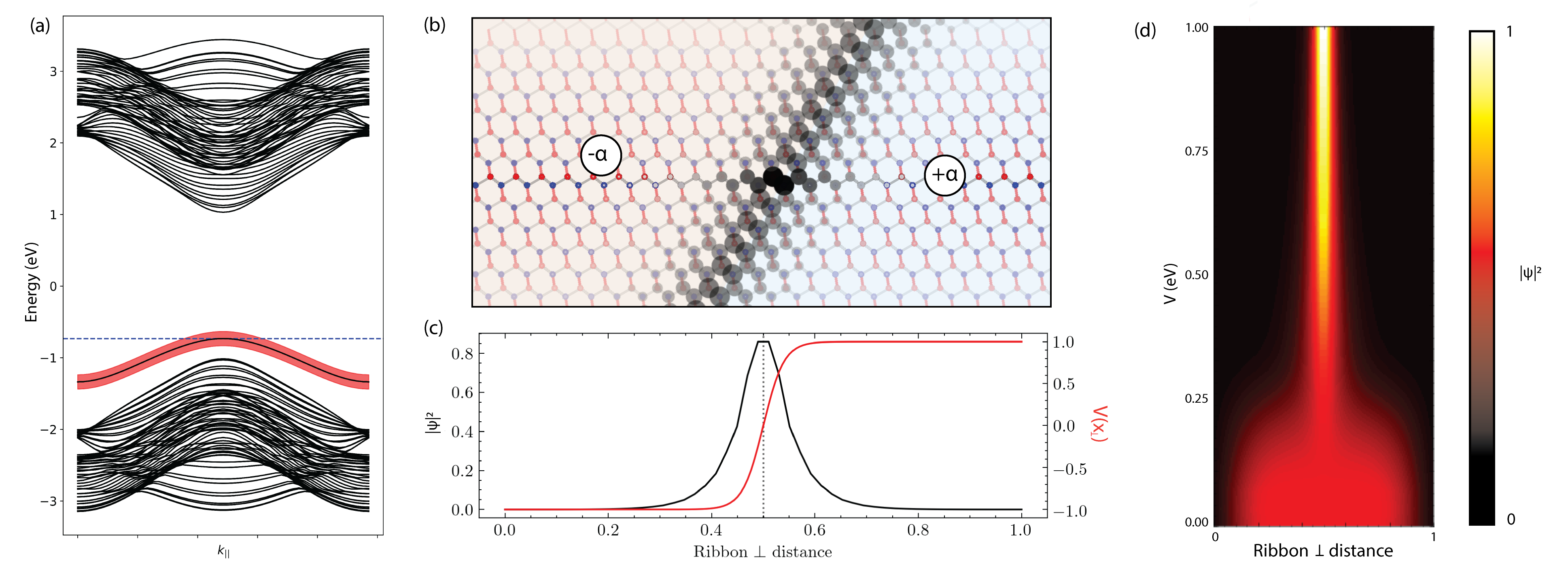} 
  \caption{ \textit{(a) } Band structure of the nano-wire when $\alpha=0.5 eV$ with the domain wall state marked in red \textit{(b)} Real space distribution of the domain wall wavefunction \textit{(c)} Gate potential profile and $\abs*{\Psi}^2$ of domain wall state. \textit{(d)} $\abs*{\Psi(\alpha,x_{\perp})}^2$ showing the localization \label{fig:1dwire-soliton}}. 
\end{figure*}

\autoref{fig:SSH-soliton} shows the eigenvalue spectrum of the system.
The states colored red are the ones localized at the origin
where the switch occurs between the two polarities. 
Since the system has lost its periodicity because of the step function, we treat the system as a large but finite number of sites.  Whether
it is trivial or not depends on the sign of $\delta t$. We consider a system with $\delta t > 0$, so that there is indeed a gap as in the SSH. In the non-trivial limit we would have an edge state which is not essential for the following discussion and hence is avoided by adding an adatom to the edges as shown in
\autoref{fig:SSH-soliton}. For the effect we consider, it is not important
whether the SSH is in the trivial or non-trivial regime as
long as there is a gap. Indeed the effect is related to the potential
shifts on the atoms rather than on the alternation of strong and weak bonds.
Alternatively, we may also envision a periodic Kronig-Penney (KP)
type
model with alternating wells and barriers. 
Starting from a finite chain, what the gate voltage
does is to flip the wells into barriers in the right half of the system,
which creates a \textit{well} of double width at the center as two atoms
with on-site shift $-\alpha$ and $-\alpha$ become adjacent at the boundary
between the two regions. For the opposite gate voltage, $+\alpha$ occurs
at both center atoms, or in the KP model, a \textit{barrier} of double width occurs.

For $\alpha=0$ (shown in middle of the top plot), one can see that the spectrum is gapped because of the interaction anisotropy from $\delta t\neq0$
but the energy states are all having comparable contribution at the
origin. In other words, the states are  extended and not localized at the domain boundary.  Next, applying a negative bias $-\alpha$ to the atoms next
to the origin, as illustrated at the bottom, we see that states localized
at the origin (indicated in red) form below and above the occupied bands.
They split farther and farther apart from the continuum of levels
as the magnitude of the bias $|\alpha|$ is increased. 

For  $\alpha=0$ as in the case of SSH, we have bonding valence and anti-bonding conduction bands in the lattice with respect to the strong bond.
Within the valence band,  the weak bonds will
form the most bonding combination of the bond-centered Wannier  orbitals
(same sign on each site) 
at the bottom of the valence band and  the most antibonding arrangement,
(alternating signs) at the top of the band. 
But with the $-\alpha$ on-site shift, at the domain boundary we create a pair of local bonding and anti-bonding states from the domain boundary adjacent
atoms. Thus
a state with even stronger bonding character forms below and another
one with more antibonding character above the VBM. 
In the KP model, one may think of the lowest state in each well as corresponding
to a strong bond of the SSH model  and these bond orbitals are now weakly coupled by tunneling through the barriers and forming a band. 
In this context, one might reason that
the center well being wider will have a lower level and hence when it interacts
with the other wells, a new level will appear below the band.  But there
will now also be a second antisymmetric band in the central well of double width
and this one will occur slightly above the band but well below the CBM
formed by broadening the antisymmetric states of the narrow wells into a
band.
The electron count does not change, so this level becomes the highest
occupied level of the system and is fully occupied.  Thus,
one  can see that we have essentially created a potential domain wall
hosting a localized state at the origin, as shown in the bottom part of
\autoref{fig:SSH-soliton}. If the system is slightly doped p-type,
say by removing one electron from the whole system, 
the split-off top level localized at the center will become partially filled,
or in other words, a hole becomes preferentially trapped at this site. 

For opposite polarity, where the two central atoms experience an upward
potential shift $+\alpha$ or a \textit{barrier}, we can see in the right  panels
of energy levels, that now states localized on the origin appear above and
below the conduction band. In the KP model one may think of this
bound state in the gap as resulting
from having a weaker tunneling trough the double well at the center. 
The highest occupied band stays at the top of the valence band but a
bound state localized near the domain wall now appears below the conduction
band. So, for slight n-type doping the electron will become localized
at the boundary.

We may call this domain wall boundary localized state a soliton state.
In fact, the boundary could be moved in principle by shifting the
region where positive and negative bias is applied, which would move
the localized state around like a solitary wave. Note, however,
that it does not originate from having two strong bonds or two weak bonds
next to each other as in the SSH soliton like states but rather
by having two adjacent atoms with downward or upward potential
(or mass shift).  

Another interesting perspective for the occurrence of this state can be deduced from topology. Breaking chiral symmetry breaks the notion of topological protection for the system, but still guarantees one localized occupied and one unoccupied edge mode (as long as SSH non triviality is satisfied)\cite{PhysRevB.83.125109}. As mentioned before, we essentially have created two copies of non-chiral SSH with one being the mirror symmetric reflection of the other (\autoref{fig:SSH-soliton} lowest panel). Thus, at the origin, the ends of these two chains meet which then topologically guarantees us a localized state. The left system consists of a lattice with $+\alpha \sigma_z$ while right $-\alpha \sigma_z$ and thus there is a domain boundary between the two ionic lattices creating a protected bound state.

Now moving on to our actual system, we can follow the same procedure to create this domain wall state. But since we have the added dimension $k_{\parallel}$, we have a dispersive band at the 1 dimensional domain wall.
To simulate a realistic device, we use $V_{gate}=\tan{(wx)}$ as shown in \autoref{fig:1dwire-soliton}(c). This essentially ensures that the domain wall is smooth and the smoothness factor is controlled by $w$.\autoref{fig:1dwire-soliton}(a)shows the band structure of this system (for $\alpha=0.6$  eV). Again, as in the case of SSH, we have added adatoms at the edges of the nanoribbon
to remove  the outer edge modes. Alternatively, one could use two semi-infinite
systems and use a Green's function technique. 
Marked in red is the localized domain wall band and its localization
is plotted as $\abs{\psi}^2$ by the gray scale
in \autoref{fig:1dwire-soliton}(b).
To numerically check the localization we also look at the $\int dk \abs{\psi(k,x)}^2$ which is exponentially localized normal to the domain boundary as shown by the applied potential profile. We also show the degree  of localization
as function of applied gate voltage in  \autoref{fig:1dwire-soliton}(d).
Clearly the higher the gate voltage, the narrower the soliton like bound
state becomes. 
Strictly speaking what we mean by gate voltage here is the applied on-site
terms on the Sb atoms. How these are related to an actual applied voltage
would depend on the technical implementation of the
device and the screening or self-consistent potential in the
Sb layer. Modeling that in detail is beyond the scope of this
qualitative paper which focuses on the concept.

These domain wall states remain occupied/unoccupied and become the highest valence/ lowest conduction bands depending  on the sign of $\alpha$ just like in the  SSH toy model. Thus by electron or hole doping the system (possibly by global gating), one could achieve localized conduction because electrons or holes
would thermalize to these domain boundary bound states. 
Interestingly, since the polarity of the channel dependents on the sign of applied voltage, one can turn a hole doped system conducting and insulating along this channel by simple reversing the global gate polarity.

Although we have here discussed the principle of the creation of a 1D wire
only, it doesn't take much imagination to see that one could apply such
opposite voltages in patches creating 1D wires at their boundaries.
By creating a pixelated gate on top and below the Sb layer, one could
imagine just changing the pattern where the voltages are applied and
hence where the wires are created by which pixels are addressed.
This is in principle a design for a reconfigurable network or circuit
of connecting wires as shown in \autoref{fig:illustration}(c). 
As mentioned above, this localization is guaranteed by topology
and thus defects/perturbations in the bulk part away from the domain wall would not interfere with this localization, hence creating robust 1D quantum
channels. In order to make the  1D wires at the boundaries conducting
the system needs to be doped in just the right way that free carriers,
holes or electrons will tend to be confined to the 1D localized domain
wall states.  This doping itself can conceivably also be realized
by a global gating of the system rather than by adding chemical dopant atoms,
whereby the overall carrier density in the system would also be controllable. 
We emphasize that domain walls of this type can be made along
several directions in the 2D Sb. Their origin depends on breaking the
inversion symmetry and is thus not confined to specific edges, such
as zigzag or arm-chair edges at which the domains meet.

\paragraph*{Summary:}
In this paper we have described two device concepts related to
the inversion symmetry breaking in monolayer Sb. Since the inversion
symmetry leads to  topologically protected edge states, breaking inversion 
does not remove the edge states but can be used to move the edge states
in energy in a
controlled way and hence their electron occupation can be controlled.
Because of the buckled nature of monolayer honeycomb Sb, this symmetry
breaking can be simply created by a gate voltage normal to the layer.
This leads to the possibility of creating a switch between
a topologically protected metallic edge state to an insulating state
protected by the large gap of the host 2D system. Furthermore in the low
temperature ballistic limit, the conductance in this 1D edge state
would be quantized to a value of $4e^2/h$ apart from a transmission
coefficient to the leads connecting to the system.  The step-like
conductivity switch would be abrupt and the off state is protected
by a large gap of these edge states toward the 2D band edges.
The bias required for switching the off-state is determined by
the band width of the dispersing edge band which follows from second
nearest neighbor interactions only and is very small compared to the gap
protecting the off state. 

Secondly,
applying the symmetry breaking of opposite polarity in patches, we
have shown that at their interface, a 1D quantum conducting wire state
could be created because of the formation of a domain wall
localized band. This process can in principle be used to create reconfigurable
conducting networks.

\paragraph{Acknowledgements -} We thank Xuan Gao, Shulei Zhang, Arvind Shankar for useful discussions. This work was supported by the U.S. Department of
  Energy-Basic Energy Sciences under grant No. DE-SC0008933.  The calculations made use of the High Performance Computing Resource in the Core Facility for Advanced Research Computing at Case Western Reserve University.

\bibliography{c3_hoti}

\end{document}